\documentstyle[11pt,aaspp4]{article}

\received{}
\accepted{}

\def\chisqnu{$\chi^2_\nu$}
\def\RXTE{{\it RXTE}}

\slugcomment{Submitted to ApJ.
Refereed version.
Draft: 981204}

\lefthead{Smith et al.}
\righthead{Multiwavelength Observations of GX 339--4}

\begin{document}
\title{Multiwavelength Observations of GX 339--4 in 1996.\\
II. Rapid X-ray Variability}

\author{I. A. Smith\altaffilmark{1}, 
E. P. Liang\altaffilmark{1}}

\altaffiltext{1}{Department of Space Physics and Astronomy, 
Rice University, MS 108, 6100 South Main Street, Houston, TX 77005-1892}

\begin{abstract}
As part of our multiwavelength campaign of GX 339--4 observations in 1996
we present the rapid X-ray variability observed July 26 using the 
{\it Rossi X-Ray Timing Explorer} (\RXTE) when the source was in a hard 
state (= soft X-ray low state).
We found that the source was extremely variable, with many bright flares.
The flares have relatively symmetric time profiles.
There are a few time intervals where the flux rises steadily and then 
drops suddenly, sometimes to a level lower than the average before 
the increase.
Hardness ratios showed that the source was slightly softer when the flux 
was brighter.
The power density spectra (PDS) were also complicated and we
found that broken power laws do not provide adequate fits to any of them.
Instead a pair of zero-centered Lorentzians gives a good general 
description of the shape of the PDS.
We found several quasi-periodic oscillations (QPO), including some that
are harmonically spaced with the most stable frequency at 0.35 Hz.
While the overall rms variability of the source was close to being constant
throughout the observation ($\sim 29$\% integrating between 0.01 and 
50 Hz), there is a small but significant change in the PDS shape with time.
More importantly, we show that the soft 2--5 keV band is more variable 
than the harder 5--10 and 10--40 keV bands, which is unusual for this source
and for other black hole candidates.
Cross correlation functions (CCF) between these bands show that the light 
curve for the 10--40 keV band lags that of the 2--5 keV band by $\sim 5$ msec.
\end{abstract}


\section{Introduction}

Most Galactic black hole candidates exhibit at least two distinct spectral 
states (see Liang \& Narayan 1997, Liang 1998, Poutanen 1998 for reviews).  
In the hard state (= soft X-ray low state) the spectrum from $\sim$ keV 
to a few hundred keV is a hard power law (photon index $1.5 \pm 0.5$) with 
an exponential cutoff.
This can be interpreted as inverse Comptonization of soft photons.
In the soft state (often, but not always, accompanied by the soft X-ray 
high state), the spectrum above $\sim 10$ keV is a steep power law 
(photon index $> 2.2$) with no detectable cutoff out to $\sim$ MeV.
This multi-state behavior is seen in both persistent sources 
(e.g. Cygnus X-1) and transient black hole X-ray novae (e.g. GRS 1009--45).
GX 339--4 is unusual in that it is a persistent source, being detected by 
X-ray telescopes most of the time, but it also has nova-like flaring states.

In 1996, we performed a series of multiwavelength observations of GX 339--4
when it was in a hard state (= soft X-ray low state).
This paper is one of a series that describes the results of that campaign.
In Paper I (\cite{smi99I}) we discuss the radio, X-ray, and gamma-ray
daily light curves and spectra obtained in 1996 July.
In Paper III (\cite{smi99III}) we discuss our Keck spectroscopy performed on 
1996 May 12 UT.
In B\"ottcher, Liang, \& Smith (1998) we use the GX 339--4 spectral data
to test our detailed self-consistent accretion disk corona models.
These papers expand significantly on our preliminary analyses 
(\cite{smi97a,smi97b}).

In this paper we focus on the rapid X-ray variability in an observation
made by the {\it Rossi X-Ray Timing Explorer} (\RXTE) 1996 July 26.
Ever since its discovery in 1973 (\cite{mar73}) it has been known that the
X-ray emission from GX 339--4 is extremely variable when it is in the 
hard state (e.g. \cite{sam79,motch83,mae84,motch85}).
This makes the source an ideal target for \RXTE, with its
unprecedented timing resolution.

In \S 2, we discuss the observing and data analysis details for the
\RXTE\ observation.

In \S 3, we describe the motivation for studying the variability in 
three bands, 2--5, 5--10, and 10--40 keV, based on the spectral components.

In \S 4, we present the light curve for our whole observation, showing
that the source was extremely variable, with many bright flares.
We focus on some representative sections of the light curve to show
the detailed evolution.

In \S 5, we present sample hardness ratio plots, showing that the
source is slightly softer when the flux is higher, though the effect
is small.

In \S 6, we present sample power density spectra (PDS), showing that they
are complicated and include harmonically spaced quasi-periodic 
oscillations (QPO).
We find that it is {\it the soft 2--5 keV band that is the most
variable}, which is unusual for black hole candidates.

In \S 7, we present sample cross correlation functions (CCF) to show that
the light curve for the 10--40 keV band lags that
of the 2--5 keV band by $\sim 5$ msec.

In \S 8 we give a brief discussion of some of the results.

\section{Observation and Data Analysis}

The \RXTE\ pointed observation was made as a Target of Opportunity 
1996 July 26 UT (MJD 50290).
A detailed discussion of the observation and data analysis is given in 
Smith et al. (1999), where we show the spectrum obtained from the Proportional
Counter Array (PCA) and the High Energy X-ray Timing Experiment (HEXTE).
Only the data from the PCA are used in the timing analysis.

The PCA consists of five Xe proportional counters with a total effective 
area of about 6500 cm$^2$ (\cite{jah96}).
It has a $\sim 1\arcdeg$ field of view, but no other bright X-ray sources 
were in the GX 339--4 field of view.
The PCA does not make separate background measurements.
Instead we used version 1.5 of the background estimator program.

Data were only used when the spacecraft was not passing through the South 
Atlantic Anomaly and when the source was observed at elevations $> 10\arcdeg$ 
above the Earth's limb and the pointing was offset $< 0.02\arcdeg$ from the 
source.
The observation was short, with good data starting 18:20:32.625 and 
ending 20:14:48.497 UT.
Because of the above constraints, valid data were only available in four
segments.
The total on-source exposure time was 3424 sec (uncorrected for dead time).

We used two data sets to generate the results presented here.
The Standard 1 production data has 0.125 sec resolution with the data
combined over all energy channels and combined over all five Proportional 
Counter Units (PCU).
Most of the results were generated from the GoodXenon data that have a 
time resolution of 0.9537 $\mu$sec and the full 256 energy channel range.
We combined all five PCU for all of the GoodXenon results.
The background contributes $\sim 13$\% of the on-source count rate when using
all the layers.
We used the \RXTE\ tasks in FTOOLS version 4.0 to extract and analyze
the data.

\section{Energy Bands}

We generated the light curves and PDS using all the energy channels combined.
We also did the same and generated the CCF using the 2--5, 5--10, and 10--40
keV energy bands (40 keV is chosen to be well below the energy where the 
background dominates at $\sim 70$ keV).
Besides having approximately the same average count rate, these three bands
were motivated by our spectroscopy using these data (\cite{smi99I}).
There we used simple power law times exponential (PLE) and Sunyaev-Titarchuk 
function (ST; \cite{suny80}) fits to the \RXTE\ data above 15 keV.
We found that two extra components were required to fit the spectrum at
lower energies:

\noindent
(1) A soft component that peaks at $\sim 2.5$ keV.

\noindent
(2) A {\it broad} emission centered on $\sim 6.4$ keV, that may be an iron 
line feature that has been broadened by orbital Doppler motions and/or
scattering off a hot medium.

\noindent
Our timing analyses using the 2--5, 5--10, and 10--40 keV bands therefore
aim to highlight the behavior of the soft, iron line, and hard model
components respectively.

However, it is important to note that the spectra are very hard.
In the ST spectral fit examples shown in Smith et al. (1999), the hard 
component dominates the entire \RXTE\ spectrum, even down to 2 keV, while 
in the PLE example, the hard and soft components cross at 4 keV.
For the various model fits, we find that the hard component 
is responsible for 95--100\% of the 10--40 keV flux.
Thus the variability analysis on the 10--40 keV band gives a clean
measurement of the variability of the hard component.
The hard component is also responsible for 45--75\% of the 2--5 keV flux.
{\it This means that the contribution of the soft component to the 
variability is significantly diluted, making it harder to discern its effect
in the 2--5 keV band.}
The hard component is also responsible for 70--90\% of the 5--10 keV flux.
The iron line component only contributes $\sim 10$\% of the 5--10 keV 
flux, and we are unable to isolate its effect.
Instead, the 5--10 keV band, where the soft component begins to have 
an impact on the variability, should be considered as an intermediary between
the other two bands.

\section{Light Curves}

Figure 1 shows the complete background subtracted light curves in the 
three bands using 0.125 sec binning.
Good data are present in four main segments separated by long gaps.
We will denote these as segments 1--4, in the order they were observed.
Figure 2 shows selected portions of the background subtracted 
light curve integrated over all PCA energies using 0.125 sec binning.
Figure 3 shows some intervals with the higher time resolution of 0.01 sec.
We find the following results:

\noindent
(1)
The source was extremely variable with a complicated light curve,
although the average fluxes in each band stayed roughly constant throughout 
the observation.

\noindent
(2)
Numerous bright flares with fluxes many times above the mean are evident,
such as Figure 2(e) that is blown up in Figure 3(c).
The bright flares last a few seconds.
Their shape is relatively triangular or has an exponential rise and decay.
Their time profiles are approximately symmetric, usually with slightly
faster decays than rises.

\noindent
(3)
There are many smaller flares that last from tens to hundreds of milliseconds.
We do not see any very short very bright spikes.

\noindent
(4)
There are a few time intervals where the flux rises steadily and then drops
suddenly, for example Figures 2(b), (g), and (h).
Sometimes the flux drops to a level lower than the average before the 
increase.
As shown in Figures 3(a), (d), and (e), these drops are resolved when 
using higher time resolution, though the drop can still be very steep.

\noindent
(5)
Broader brightenings and dimmings that last $\sim 10$ sec are seen, such
as in Figure 2(a).

\noindent
(6)
There are broad U-shaped dips where the flux drops below the average,
such as in Figure 2(d) which is expanded in Figure 3(b).

\noindent
(7)
The light curves in the three energy bands are almost identical.
There is a very good correspondence in the flaring behavior in the different 
bands, with a tendency for the flares to be slightly softer, though this is 
not always the case.

\section{Hardness Ratios}

Figure 4 shows the hardness ratio versus intensity plots for the 
three energy bands using 16 second binning.
There is a weak trend, with the source being slightly softer when the 
flux is higher.
This is consistent with the light curves, where most often the flares were
slightly softer.
The most significant correlation is in the 2--5 vs 10--40 keV bands, as might
be expected since these bands show the cleanest separation of the soft and
hard components.

The 16 second binning integrates over all the finer rapid variability.
Unfortunately, when we do the hardness ratio analysis on shorter time scales, 
the error bars become too large to determine any trends.

We do not see any hardness ratio correlations with time during the whole run.
The fact that the hardness ratios show little dependence on time or
brightness indicates that the spectral shape does not change markedly.
Thus the average spectrum shown in Smith et al. (1999) is a good
representative shape, though it should be noted that the amplitude 
varies dramatically on short time scales.

\section{Power Density Spectra}

We generated the power density spectra (PDS) to look for 
quasi-periodic oscillations (QPO) in the source.

The original bin size of the data is 0.954 $\mu$sec ($1/2^{20}$ sec).
In all the PDS shown, we have summed the data into 10.0 msec bins.
The power spectra were generated for intervals of 81.92 sec and the
results were then averaged.
We always truncated the four segments of data to ensure that only
completely filled intervals were used.
The PDS frequency bins were binned logarithmically in the figures.

Given the large gaps between the segments of good data, we performed the 
PDS analyses on the four segments separately.
This avoids spurious effects due to data gaps, and allows us to look for
variations in the PDS behavior with time.
Figure 1 shows that the first two segments have longer stretches of data,
and thus these have the better statistics in the timing analyses.

In all cases, we have checked that the power has a white noise spectrum
at high frequencies, as expected from counting statistics.
In all the PDS plots shown, we have subtracted this white noise component.
The plots are normalized so that their integral gives the squared rms
fractional variability, i.e. the power spectrum is in units of rms$^2$/Hz.

Figure 5 shows the PDS for the four data segments using all PCA energies.
The model fit parameters are given in Table 1.
Given the complicated light curve, it is not surprising that the 
PDS are also complicated.
Several QPOs are evident, though we do not see any in the kHz region.

\placetable{pdstable}

A common fit to the PDS shape is a broken power law:
$P(\nu) = n~(\nu \leq \nu_0),~P(\nu) = n (\nu/\nu_0)^k~(\nu > \nu_0)$
where $n$ is the amplitude and $\nu_0$ the break frequency.
{\it We found that this broken power law does not fit any of our results.}
For the ``best fits'' to segments 1, 2, 3, and 4, the reduced 
\chisqnu = 3.4, 4.0, 1.5, and 1.8 respectively.
The probabilities that random sets of data points would give values of
\chisqnu\ as large or larger than these are
$Q = 6.2 \times 10^{-25}, 1.1 \times 10^{-32}, 
3.2 \times 10^{-3}$, and $6.5 \times 10^{-6}$.

A pair of zero-centered Lorentzians gives a much better description of
the general PDS shape.
This corresponds to an exponential shot model with shots of two distinct
time constants (\cite{gro94,smi97c}).
The Lorentzian has distribution
$$P(\nu) = {LN \over {1 + [2(\nu - LC)/LW]^2}}$$
where $LC$ is the centroid frequency (in Hz),
$LN$ is the normalization at $\nu=LC$ (in rms$^2$/Hz),
$LW$ is the full width at half maximum, and the 
area (integral) $I=\pi(LW)(LN)/2$ (in rms$^2$).
For the best fits to the four segments, the reduced \chisqnu = 1.65, 1.92, 
0.88, and 0.99 respectively for 105 degrees of freedom, giving
$Q = 3.1 \times 10^{-5}, 4.0 \times 10^{-8}, 0.81$, and 0.52.
It is apparent that the PDS are more complicated than a simple pair
of zero-centered Lorentzians, though this is only significantly the
case for the first two segments.
A pair of zero-centered Lorentzians is an adequate fit to 
segments 3 and 4, where the statistics were poorer: 
these fits are shown in Figures 5(c) and 5(d).

For segment 1 with just two zero-centered Lorentzians, the \chisqnu = 1.65
and $Q = 3.1 \times 10^{-5}$.
Figure 5(a) shows that apparently harmonically spaced QPO features around 
$\sim 0.1$ Hz are most noticeable in this segment.
We started by adding just one Lorentzian line.
Centering this at 0.35, 0.175, or 0.0875 Hz improved the fit to 
\chisqnu $\sim 1.5$, $Q \sim 6 \times 10^{-4}$ in each case.
Thus one line is an improvement, but is inadequate on its own.
Adding two harmonically spaced Lorentzian lines centered at 0.35 and 0.175 Hz 
improves the fit to \chisqnu = 1.30, $Q = 0.023$.
Now adding a third QPO at 0.0875 Hz slightly improves the fit to 
\chisqnu = 1.26, $Q=0.041$.
Although a fourth QPO feature at 0.044 Hz appears to be present, we are
unable to prove this significantly: the interval length of 81.92 sec
prevents getting results at low frequencies, while using longer intervals
results in fewer of them and the error bars increase.
Allowing the centroid frequencies of the QPO peaks to vary
does not improve the fits (given
the reduction in the number of degrees of freedom), and so it seems
likely that the lines are indeed harmonically spaced.
Other possible QPO at higher frequencies are also apparent.
Indeed, adding a narrow Lorentzian line at 6.1 Hz improves the fit
to \chisqnu = 1.14, $Q = 0.17$.
However, we found that it was not possible to determine a confidence 
region for this line, because the program instead found a smaller
\chisqnu\ with a very broad peaked noise feature.
In Figure 5(a) we show the fit that just uses a pair of zero-centered
Lorentzians plus three harmonically spaced Lorentzian QPO.

Figure 5(b) shows that segment 2 has less pronounced QPO features below 1 Hz 
than segment 1, but it has a much more complicated PDS in the 1--20 Hz region.
For segment 2 with just two zero-centered Lorentzians, the \chisqnu = 1.92
$Q = 4.0 \times 10^{-8}$.
Adding a Lorentzian line centered at 0.35 Hz improves the fit 
significantly to \chisqnu = 1.46, $Q= 1.6 \times 10^{-3}$.
Neither allowing the centroid frequency to vary
nor adding other lines below 0.3 Hz
improved the fit (given the reduction in the number of degrees of freedom).
This shows that the 0.35 Hz line was the most stable of the harmonic lines 
in segment 1.
As in segment 1, adding a very broad peaked noise feature at 3.6 Hz greatly
improves the fit to \chisqnu = 1.11, $Q= 0.22$.
Figure 5(b) shows this model that uses a pair of zero-centered
Lorentzians, one Lorentzian QPO, and one broad peaked noise feature.

Integrating, the fractional rms variability between 0.01 and 50 Hz is
29\% for segment 1, 30\% for segment 2,
26\% for segment 3, and 28\% for segment 4.
As expected, the fractional rms variabilities are very large.
The overall variability remains approximately the same throughout the 
whole observation, as is apparent from the similar light curves of the four
segments in Figure 1.
However, the detailed shape of the PDS does change with time.
Using the three QPO model from segment 1 for segment 2 gives \chisqnu = 1.69, 
$Q = 9.7 \times 10^{-6}$, showing this is not an acceptable fit.
Furthermore, splitting segments 1 and 2 into shorter sections, there is weak 
evidence that the 0.0875 Hz line was narrower in the first half of segment 1
and that the broad peaked noise at 3.6 Hz was most prominent in the last 
quarter of segment 2.
For segment 1 the fractional rms amplitude is 
19\% for the 0.35 Hz QPO, and 6\% for the other two QPO.
For segment 2, the fractional rms amplitude is 
22\% for the 0.35 Hz QPO, and 9\% for the 3.6 Hz peaked noise.

As for the light curves, we generated the PDS using the 2--5, 5--10, 
and 10--40 keV bands.
We again summed the data into 10.0 msec bins and generated the PDS for 
intervals of 81.92 sec that were then averaged.
Given the lower count rates in the three bands, larger logarithmic 
frequency bins have been used than in Figure 5.
Figure 6 shows the results for segment 1.
We find that the model fit used in Figure 5(a) (where all the PCA energies were
combined) also gives good fits to the individual bands, provided the whole
model is multiplied by a constant: this constant is 1.4 for the 2--5 keV
band, 1.2 for the 5--10 keV band, and 2/3 for the 10--40 keV band.
The fit results are given in Table 2.
We get the same results for segments 2, 3, and 4: multiplying the 
corresponding model in Figures 5(b), 5(c), and 5(d) by 1.4, 1.2, and 
2/3 gives good fits to the observed PDS in the 2--5, 5--10, and 10--40 keV 
bands respectively.
Integrating, the fractional rms variability between 0.01 and 50 Hz is
35\% for 2--5 keV, 32\% for 5--10 keV, and 24\% for 10--40 keV.
{\it This means that the variability is greatest in the 2--5 keV band
and smallest in the 10--40 keV band: it is very unusual for the soft
band to be the most variable one in galactic black hole sources.}
This result is even more striking bearing in mind that the hard component
gives a large if not the dominant contribution to the 2--5 keV band, 
strongly diluting the effect of the soft component.
The energy dependence of the PDS agrees with the light curves, where it was 
apparent that the bright flares are most often softer.
The fact that the PDS shapes are consistent with being the same for the 
three bands also agrees with the light curves, where there was a good 
tracking between the three bands.
This shows that there is no obvious energy dependence at different 
frequencies.

\placetable{pdstable}

\section{Correlation Functions}

We generated the auto correlation functions (ACF) for the four segments using
all the PCA energy channels.
The ACF are complex, and are not amenable to simple fitting.
A single exponential gives a very poor fit, even using a small range of time
delays for the fitting.
The characteristic decay time scale (coherence time) is $\sim 0.4$ sec for
all four segments.
Since the PDS contain the same information in a more useful form, we do 
not discuss the ACF results further.

A detailed frequency dependent phase lag analysis and interpretation
using Compton scattering models is underway and will be presented elsewhere.
In this paper we simply present the cross correlation functions (CCF) of the
light curves in the 2--5, 5--10, and 10--40 keV bands.

As before, we generated the CCF for the four segments separately.
Figure 7 shows sample CCF for segment 1.
We have removed Poisson noise from the figures.
The results are shown for the 2--5 keV band compared to the 10--40 keV band.
We use three different bin times, 0.1, 0.01, and 0.0025 sec to investigate
delays on different time scales.
Positive time delays imply that the 10--40 keV light curve lags the 
2--5 keV one.

The figure shows that for the longer time delays the CCF is relatively 
symmetric.
This is not surprising, since there was no dramatic hardening or 
softening during our run.
There is a broad bump at $\pm 2$ sec that is also seen in the ACF.
For shorter time delays the CCF is not symmetric, and there is a small offset 
from 0.
{\it The 10--40 keV light curve lags that of the 2--5 keV band by $\sim 5$ 
milliseconds.}
We get similar CCF using the 2--5 and 5--10 keV bands, although they are
more symmetric and there is no significant offset from 0.
We also get similar CCF using the 5--10 and 10--40 keV, but with an
offset of $\sim 2.5$ ms (the 10--40 keV light curve lags that 
of the 5--10 keV band).
As we noted in \S3, the 5--10 keV band should be considered as an 
intermediary between the other two bands, where the soft model component 
begins to have an impact on the variability, so getting an offset that is
in between the other two cases was expected.

We find essentially the same results for the CCF generated from the other 
three segments, in particular in the values of the offsets.
This gives us confidence that there is a real delay between the hard and
soft photons.
However, we caution that the error bars are relatively large because the 
number of counts is relatively low when we are trying to extract the time 
delays on such short time scales.

\section{Discussion}

Our \RXTE\ observation showed that GX 339--4 was extremely variable 
on 1996 July 26.
The flaring light curve and the shape of the PDS and the QPO are quite
similar to previous {\it Ginga} observations (Figure 1 of \cite{miy92} and
Figure 3.5 of \cite{tan95}).
This indicates that the results presented here are probably fairly typical 
of the source in the low (hard) state.
However, we caution that we only have a small stretch of data, and our
results need to be compared and contrasted with those from other \RXTE\ 
observations.

In shot models, it is often assumed that the shots have a sharp rise and an
exponential decay or vice versa (e.g. \cite{bel97}), although the PDS 
predicted by the shot models are generally insensitive to the true shape 
of the flare.
A sharp rise or fall is not seen in most of our flares, that instead are 
relatively symmetric.
The shots in GX 339--4 therefore appear to be similar to those 
in Cygnus X-1 (\cite{nmk94}).
We do see a few places in the light curve where the flux slowly rises and
then has a rapid decay, sometimes to a level lower than the average flux.
This may indicate that a denser ring of material fell onto the compact
object, partially emptying the accretion disk.

We found that a broken power law does not fit any of our PDS.
Instead a pair of zero-centered Lorentzians gives a good general 
description of their shape.
The PDS for {\it EXOSAT} observations of GX 339--4 were fit by a single
zero-centered Lorentzian by M\'endez \& Van der Klis (1997).
Their Figure 3 plotted the fractional rms at the break frequency versus
the break frequency for many black hole candidates.
While it is not straightforward to add our results to this plot because
our PDS have a very complex shape, roughly the break frequency is at 0.2 Hz
and the fractional rms at the break frequency is $\sim 33$\%.
This is consistent with the other observations in the upper branch of their
figure that come from the low state observations of the sources.

We found several QPOs in our observation.
The ones around 0.1 Hz are harmonically spaced in the first segment of data.
The QPO at 0.35 Hz is also present in the second segment of data.
QPOs have been seen in several previous observations of GX 339--4, both 
in the low and very high states.
These results are summarized in Table 6.2 of Van der Klis (1995).
We also found evidence for QPO in {\it ASCA} observations of this source
(\cite{dob97}).
Harmonically spaced QPOs similar to ours were seen at 0.05 and 0.1 Hz 
in the X-ray low state by Motch et al. (1983).
Harmonically spaced QPOs were also seen by Miyamoto et al. (1991) and
Belloni et al. (1997), though these are in the very high state, and 
the QPO were at higher frequencies than those found here.
It may be interesting to note that a couple of months prior to our \RXTE\
observation (in 1996 April) a long-lived 0.064 Hz QPO 
was seen in the optical emission from GX 339--4 (\cite{steim97}).

While the overall rms variability of the source was close to being constant
throughout the observation ($\sim 29$\% integrating between 0.01 and 
50 Hz for all PCA energies), there is a small but significant change 
in the PDS shape with time.
Our light curves show that the $\sim 0.1$ Hz QPO in segment 1 may be related
to the broad $\sim 10$ sec brightenings and dimmings (e.g. Figure 2(a));
this effect was less evident in the later segments.
The apparent structure in the PDS above 1 Hz presumably comes from the 
contributions of the individual shorter flares.
However, the overlap of many flares results in a complicated PDS.

We found that the shape of the PDS has no obvious energy dependence.
However, the amplitude does, with the soft 2--5 keV band having the
greatest variability.
This is a very surprising result.
Previous observations of GX 339--4 in the low state by Motch et al. (1983)
and Maejima et al. (1984) found that the PDS are usually insensitive to the 
energy band studied.
Normally when galactic black hole candidates do have differences, it is the 
hard band that is the more variable one (\cite{vdk95}).
Thus this needs to be further
investigated through more observations of this source.

The fact that the light curves in the different energy bands track 
closely suggests that the emission mechanisms responsible for the soft and
hard components are not independent.
This is consistent with Compton reprocessing models.
However, the fact that the soft band is the most variable one in our 
observation indicates that the region responsible for the hard component 
does not fully cover the region responsible for the soft component.

We found that the 10--40 keV light curve lags that of the 2--5 keV band by 
$\sim 5$ ms.
This is consistent with the hard component being due to Compton scattering
of the soft component.
The 5 msec delay implies a typical path length difference of 1500 km for
the hard and soft photons.
The delay of the hard X-rays in GX 339--4 resembles the results found 
for Cygnus X-1 in the low state (\cite{miy88}).

Our CCF results investigate the time delays on shorter time 
scales than in most of the previous studies of GX 339--4 in the low state.
Our results are therefore consistent with those of Maejima et al. (1984),
who saw no time delays in their CCFs that only had 187.5 ms time resolution.
Similarly, Motch et al. (1983) found that the CCF of different energy
channels were synchronized to better than 80 ms.
Miyamoto et al. (1992) used {\it Ginga} observations to show that the
4.6--9.2 keV band lagged the 1.2--4.6 keV band on a $\gtrsim 10$ ms time scale.
While they suggested the phase lag had a frequency dependence, their results
appear to be equally consistent with no frequency dependence.
Their 1.2--4.6 keV band samples the soft component slightly better than our 
2--5 keV band, that contains a large contribution from the hard component.
This may explain why the shape of our PDS and light curves show no 
significant difference between the three bands.
Miyamoto et al. (1991) also found time delays when GX 339--4 was in a very
high state: the time variations in the 2.3 -- 4.6 keV band were the most
advanced.

\acknowledgments

We thank the referee for carefully reading the manuscript and providing
useful suggestions and clarifications.
This work was supported by NASA grants NAG 5-1547 and 5-3824 at 
Rice University.

\clearpage

\begin{figure} \plotone{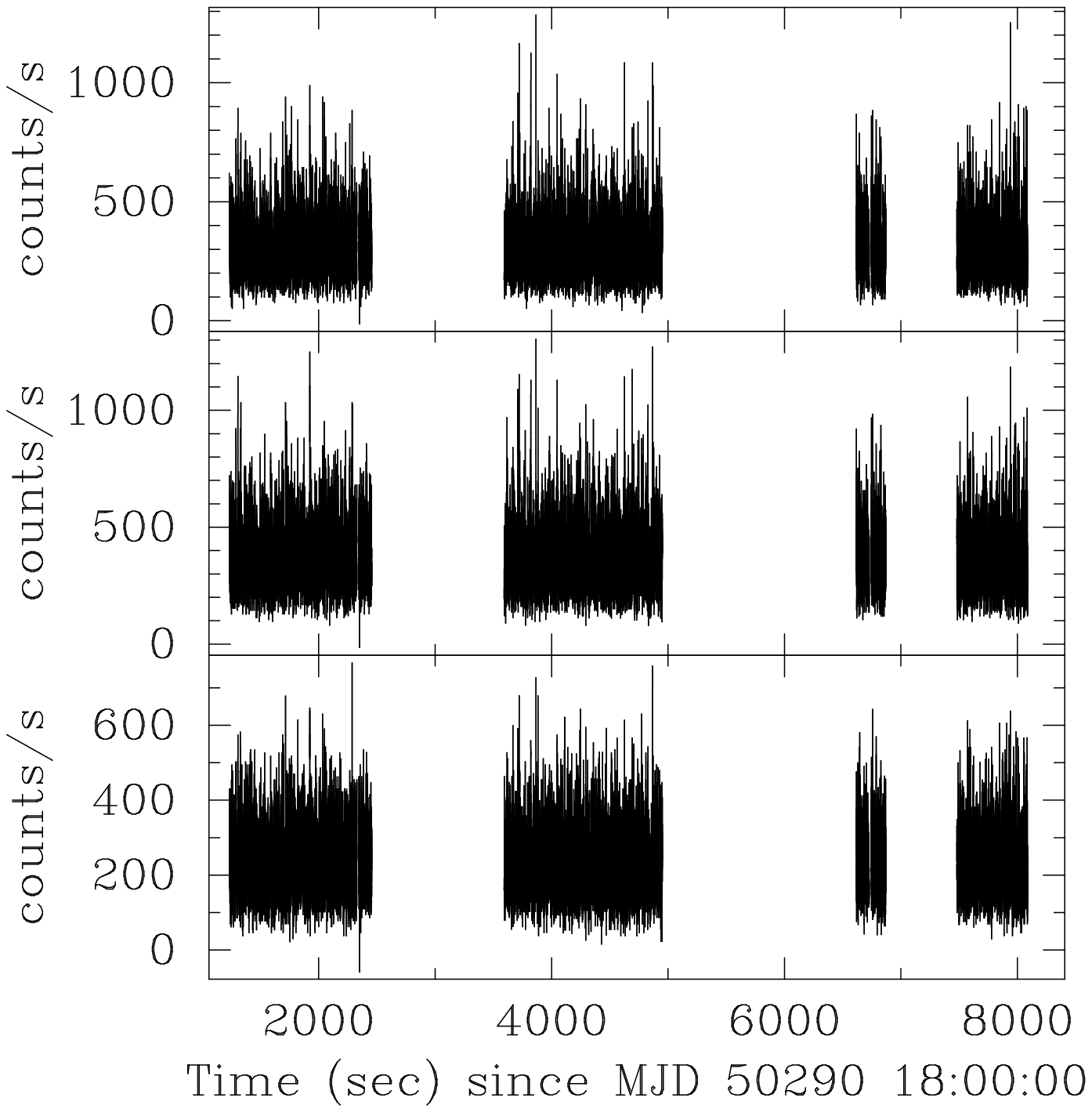} \begin{center}
\figcaption{Background subtracted light curves for the 1996 July 26 
\RXTE\ observation of GX 339--4.
The bin time is 0.125 sec.
Top: 2--5 keV.
Middle: 5--10 keV.
Bottom: 10--40 keV.}
\end{center} \end{figure}

\begin{figure} \plottwo{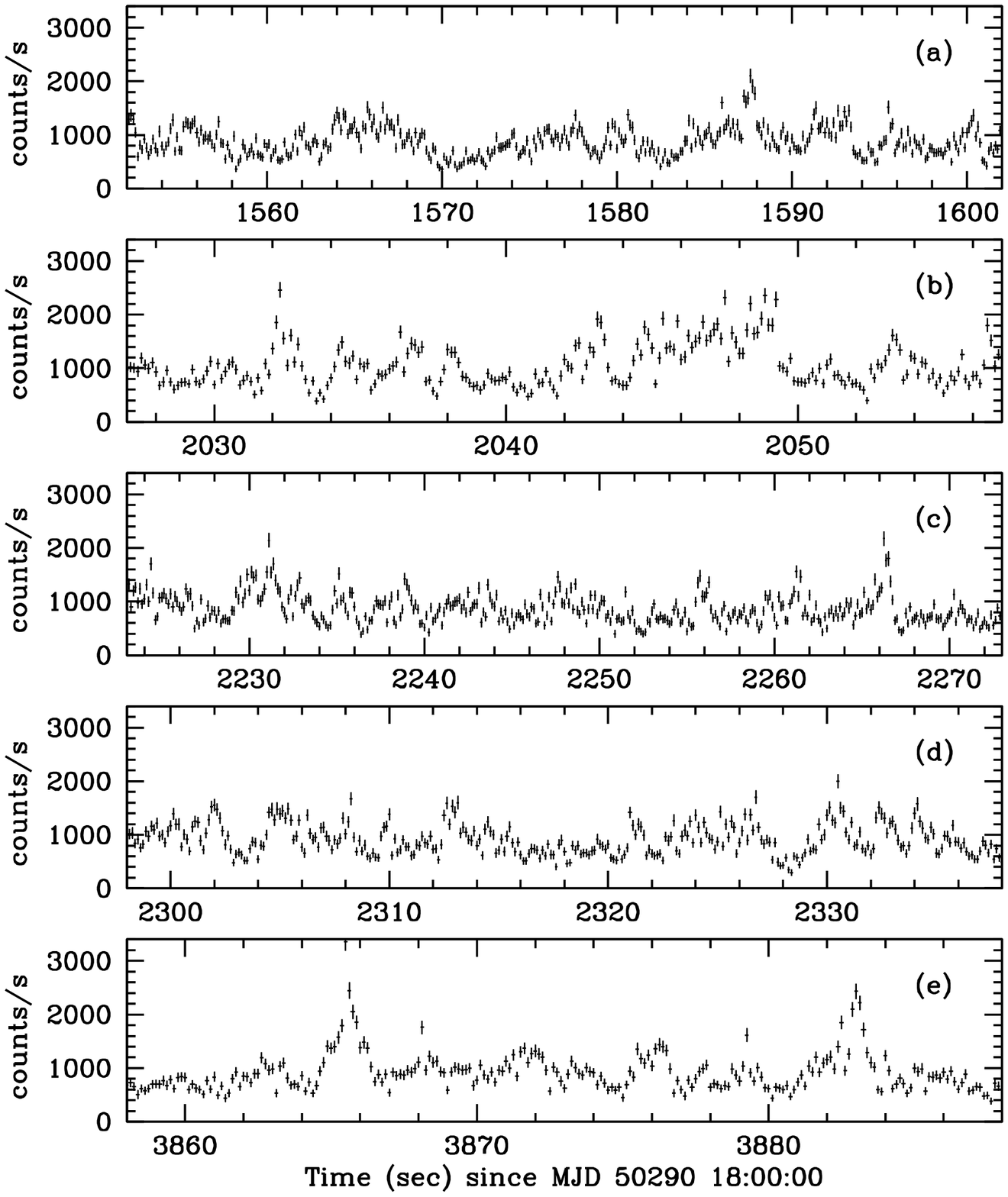}{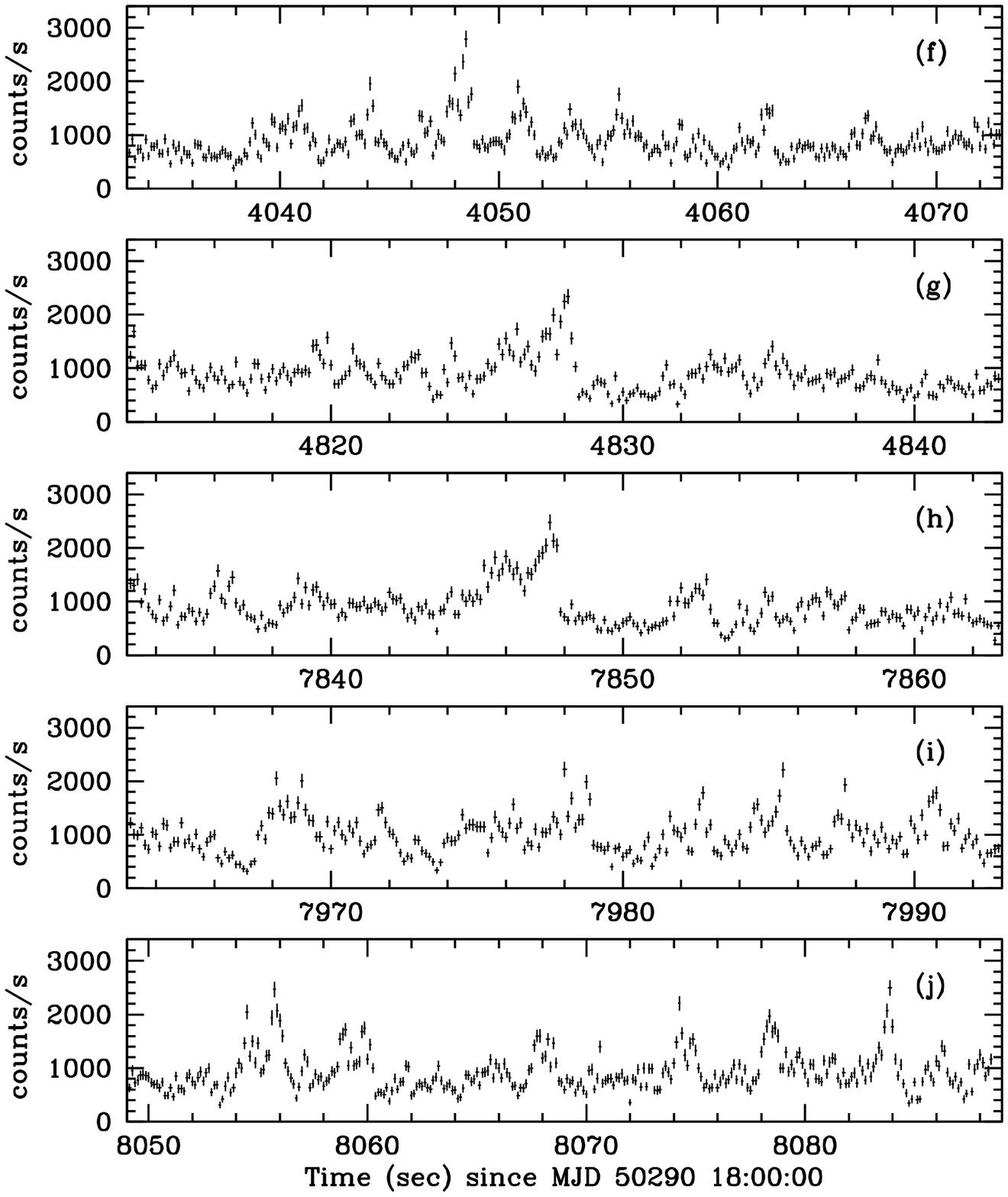} \begin{center}
\figcaption{Sample sections of the background subtracted light curve.
The bin time is 0.125 sec.
We have integrated over all PCA energies.}
\end{center} \end{figure}

\begin{figure} \plotone{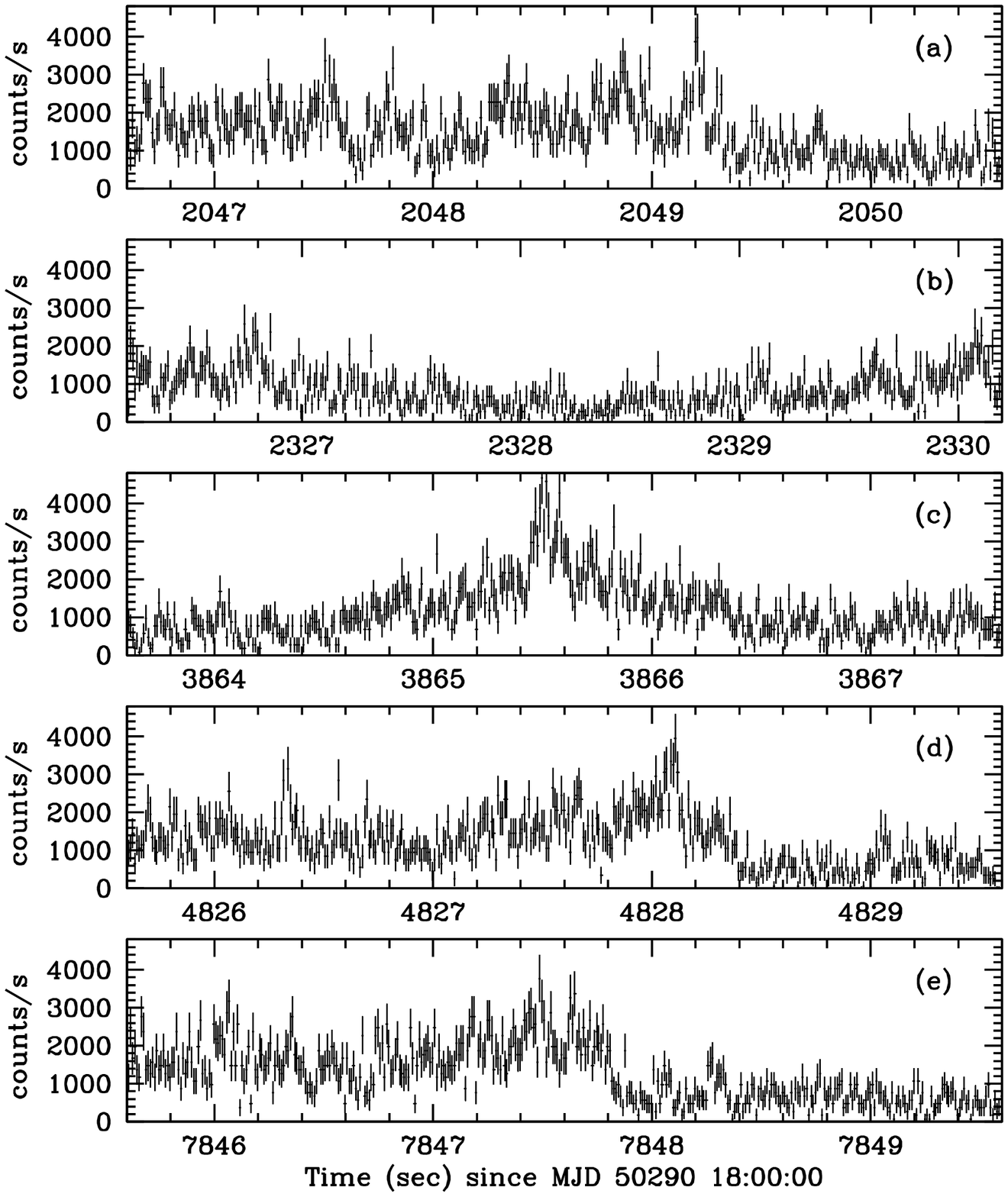} \begin{center}
\figcaption{Sample sections of the background subtracted light curve.
The bin time is 0.01 sec.
We have integrated over all PCA energies.}
\end{center} \end{figure}

\begin{figure} \plotone{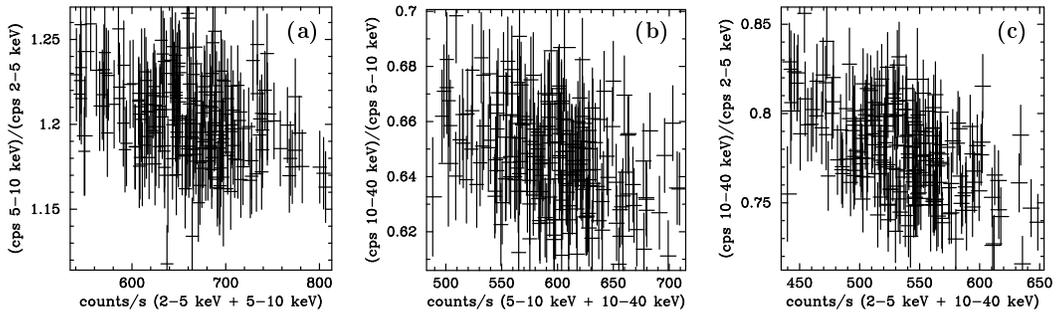} \begin{center}
\figcaption{Hardness ratios versus summed intensities.
16 sec binning has been used.
The energy bands used are
(a) 2--5 keV and 5--10 keV,
(b) 5--10 keV and 10--40 keV,
(c) 2--5 keV and 10--40 keV.}
\end{center} \end{figure}

\begin{figure} \plotone{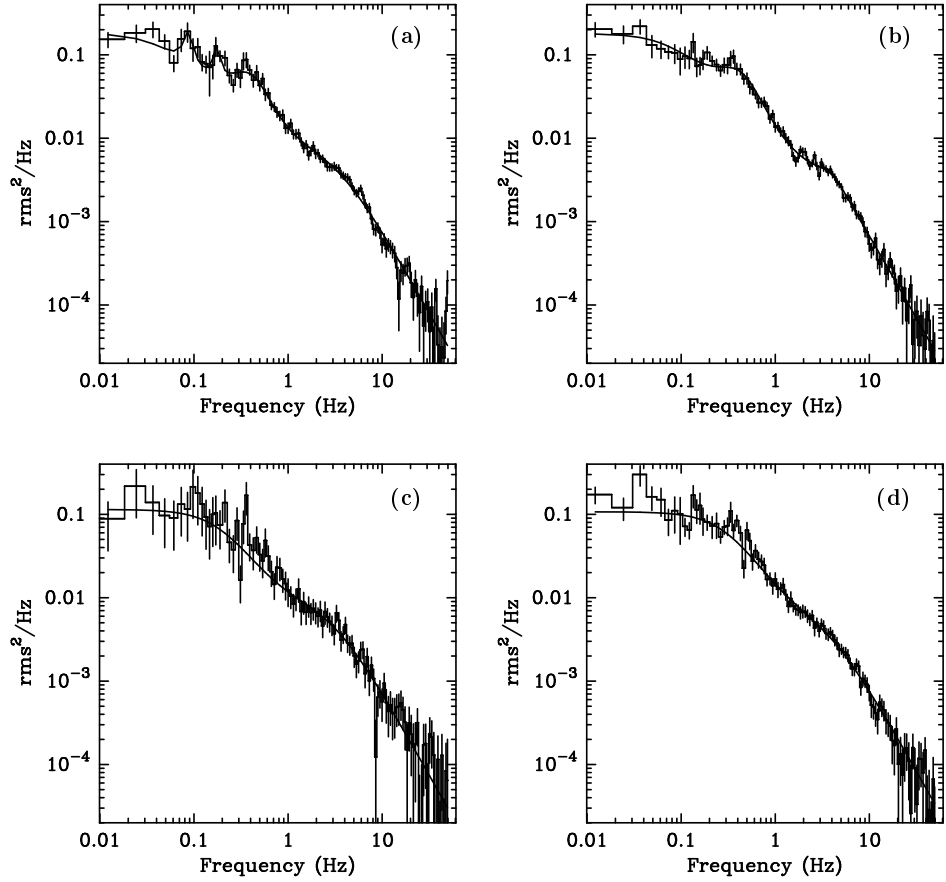} \begin{center}
\figcaption{PDS for each segment using all PCA energies.
Intervals of length 81.92 sec have been averaged.
The model fits in each case are given in Table 1.
(a) Segment 1 (14 intervals).
(b) Segment 2 (16 intervals).
(c) Segment 3 (3 intervals).
(d) Segment 4 (7 intervals).}
\end{center} \end{figure}

\begin{figure} \plotone{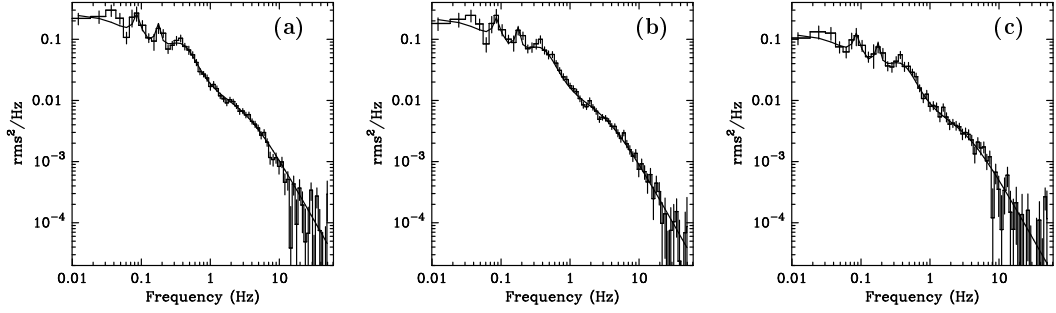} \begin{center}
\figcaption{PDS for segment 1 for different energy bands.
In each case 14 intervals of 81.92 sec have been averaged.
The model curve is identical to the one in Figure 5(a) multiplied by a
scaling factor.
The fit results are given in Table 2.
(a) 2--5 keV band (scaling factor 1.4).
(b) 5--10 keV band (scaling factor 1.2).
(c) 10--40 keV band (scaling factor 2/3).}
\end{center} \end{figure}

\begin{figure} \plotone{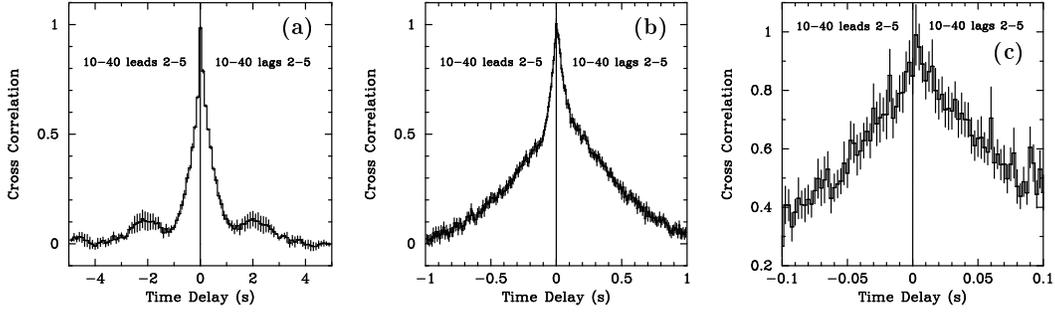} \begin{center}
\figcaption{CCF for segment 1 using different time binning.
Positive time delays imply that the 10--40 keV light curve lags the 
2--5 keV one.
(a) Using 0.1 sec time bins, 5 intervals of 204.8 sec have been averaged.
(b) Using 0.01 sec time bins, 59 intervals of 20.48 sec have been averaged.
(c) Using 0.0025 sec time bins, 239 intervals of 5.12 sec have been averaged.}
\end{center} \end{figure}

\clearpage

\begin{deluxetable}{cccccccc}
\tablewidth{0pt}
\tablecaption{PDS fits using all PCA energies \label{pdstable}}
\tablehead{
\colhead{Segment} &
\colhead{Component\tablenotemark{a}} &
\colhead{LC (Hz)} &
\colhead{LW (Hz)} &
\colhead{LN (rms$^2$/Hz)} &
\colhead{$\chi^2_\nu$} &
\colhead{DOF} &
\colhead{$Q$}}
\startdata
1 & ZCL & 0 & $0.10 \pm 0.03$ & $0.16 \pm 0.03$ & 1.26 & 99 & 0.04 \nl
 & ZCL & 0 & $6.3 \pm 0.3$ & $0.0078 \pm 0.0005$ \nl
 & QPO & 0.35 & $0.47 \pm 0.04$ & $0.050 \pm 0.004$ \nl
 & QPO & 0.175 & $0.030 \pm 0.013$ & $0.076 \pm 0.030$ \nl
 & QPO & 0.0875 & $0.022 \pm 0.012$ & $0.12 \pm 0.05$ \nl
\nl
2 & ZCL & 0 & $0.18 \pm 0.05$ & $0.15 \pm 0.03$ & 1.11 & 100 & 0.22 \nl
 & ZCL & 0 & $7.8 \pm 0.8$ & $0.0038 \pm 0.0012$ \nl
 & QPO & 0.35 & $0.57 \pm 0.06$ & $0.055 \pm 0.006$ \nl
 & PN & $3.6 \pm 0.5$ & $3.5 \pm 1.3$ & $0.0015 \pm 0.0004$ \nl
\nl
3 & ZCL & 0 & $0.41 \pm 0.10$ & $0.11 \pm 0.02$ & 0.88 & 105 & 0.81 \nl
 & ZCL & 0 & $5.8 \pm 0.6$ & $0.0083 \pm 0.0014$ \nl
\nl
4 & ZCL & 0 & $0.64 \pm 0.05$ & $0.10 \pm 0.01$ & 0.99 & 105 & 0.52 \nl
 & ZCL & 0 & $7.3 \pm 0.6$ & $0.0055 \pm 0.0007$ \nl
\enddata
\tablenotetext{a}{ZCL = zero-centered Lorenztian, QPO = quasi-periodic
oscillation, PN = peaked noise.}
\tablecomments{Errors are 68\% confidence regions for varying one 
parameter.}
\end{deluxetable}

\begin{deluxetable}{ccccc}
\tablewidth{0pt}
\tablecaption{PDS fits for segment 1 using energy bands \label{pds1etable}}
\tablehead{
\colhead{Band (keV)} &
\colhead{Scaling factor\tablenotemark{a}} &
\colhead{\chisqnu} &
\colhead{DOF} &
\colhead{$Q$}}
\startdata
2--5 & 1.4 & 1.19 & 63 & 0.14 \nl
5--10 & 1.2 & 1.00 & 63 & 0.47 \nl
10--40 & 2/3 & 1.14 & 63 & 0.21 \nl
\enddata
\tablenotetext{a}{Constant multiplying the model curve in Figure 5(a).}
\end{deluxetable}

\end{document}